\documentstyle[rotating,12pt]{crnart12}
\begin{document}
\normalsize
\begin{flushright}
{\large
BARI-TH/172-94 \\
April 1994}
\end{flushright}

\begin{center}

\vspace*{15mm}
\normalsize

{\huge {\bf Exotics Searches}}
\normalsize

\vspace*{9mm}
{\large
Michael Doser \\
\vspace*{3mm}
\it{CERN, Geneva, Switzerland} \\
\vspace*{3mm}}
\normalsize
{\large
and \\
\vspace*{3mm}
Antimo Palano \\
\vspace*{3mm}
\it{Dipartimento di Fisica dell'Universit\'a and \\
Sezione INFN, Bari, Italy} \\}
\end{center}
\normalsize

\vspace*{25mm}

\begin{center}
\par
{\large\bf Abstract} \\
\end{center}
The experimental information on the search for non $q \bar q$ mesons as
glueballs, hybrids and multiquark states is reviewed.
Candidate states which are particularly amenable to detailed study
by accumulating large samples of $J/\psi$, $\eta_c$, $\chi$ decays
at a $\tau$-charm factory are discussed.
\vspace*{60mm}

\begin{center}
{Invited talks to the $3^{rd}$ Workshop on a $\tau$-charm factory \\
Marbella, Spain, June 1-6 1993}
\end{center}
\newpage
\section*{\normalsize\bf 1. INTRODUCTION}
\vskip 8pt
The present understanding of strong interactions is that they are
described by Quantum Chromo Dynamics (QCD). This non-Abelian field
theory not only describes how quarks and antiquarks interact, but
also predicts that the gluons which are the quanta of the field will
themselves interact to form mesons. If the object formed is composed
entirely of valence gluons ($gg$ or $ggg$) the meson is called
a glueball, however
if it is composed of a mixture of valence quarks, antiquarks and
gluons (i.e. $q \bar q g$) it is called a hybrid. In addition,
$q \bar q q \bar q$ states are also predicted.
An unambiguous confirmation of these states would be an important
test of QCD and
would give fundamental
information
on the behaviour of this theory in the confinement region.
\par
However, until large dedicated computers (QCD machines) are
available, in order to compute QCD with sufficient precision,
the spectrum of the hadrons will be only known in a qualitative way.
The perturbative approach cannot, in  fact, be easily
extended to the low energy regime; the light hadron spectrum cannot be
reliably calculated, and it is even more difficult to predict dynamical
properties, such as decay widths. The spectroscopy of low-mass states
can however be accounted for, to a large degree, by QCD-inspired models.
The most complete of these, built by S. Godfrey and
N. Isgur in 1985~\cite{Isgur}, is able to describe with sufficient accuracy the
$q \bar q$ mesons spectrum
from the pion to the $\Upsilon$. This model is therefore often used in order to
test whether a new discovered resonance belongs or not to one of the
$q \bar q$ multiplets.

Although the existence of a glueball spectrum is predicted by QCD,
the extraction of reliable predictions for the masses of these states
presents an important challenge.
Recent advances in
lattice gauge theory calculations~\cite{UKQCD} are however beginning
to shed light on both ordering of states and mass scale, giving values
of 1550 $\pm$ 50 MeV and of 2270 $\pm$ 100 MeV
for the lowest-lying 0$^{++}$ and 2$^{++}$ glueballs.
While the absolute scale is still uncertain, the mass ratio
prediction is in line with previous values from various other
theoretical models (MIT bag model, potential models, QCD sum rules, flux-tube
model)~\cite{glueballpredictions}.

The mesonic decay of glueballs is determined by
their flavor SU(3) singlet nature; ignoring phase space factors, glueballs
are naively expected to couple equally to all flavors, while
arguments from perturbation theory~\cite{chansharpe}
favor a stronger coupling to strange, rather than to u- or d-quarks.
It is in any case important to note that these production and decay
characteristica rest on the assumption of pure
gluonium states, and may be considerably modified by an
admixture of $\bar{\mbox{q}}$q states.

\section*{\normalsize\bf 2. THE SEARCH FOR NON $\bar q q$ STATES}
\vskip 8pt

The experimental search for gluonium states started in 1980 with the
discovery, by Mark II~\cite{iota1} and the Crystal Ball~\cite{iota2}
experiments at SPEAR of a large "$\iota/\eta(1440)$"
signal (m=1440 MeV, $\Gamma$=50 MeV) in
the radiative $J/\psi$ decay $J/\psi \to \gamma K \bar K \pi$
(see fig.~\ref{iota}).
\begin{figure}[htp]
\vspace*{10cm}
\caption[iota]
{\small\it
\label{iota}
Observation of the $\iota/\eta(1440)$ by: a,b) MarkII and
c) Crystal Ball}
\end{figure}
The quantum numbers of this state have been determined by the Crystal
Ball experiment to be $J^{PC}=0^{-+}$.
Since the pseudoscalar nonet was already full and since the radiative $J/\psi$
decay was understood to proceed through a two gluon intermediate state
as shown in fig.~\ref{graph}a), the new resonance
was readily proposed as the first candidate for being a gluonium state.

This finding and the discovery of the $\theta(1720)$ by
the Crystal Ball~\cite{theta} in the reaction $J/\psi \to \gamma \eta \eta$,
motivated a new interest in meson spectroscopy so that the
search for gluonic mesons has been the main motivation
of light meson spectroscopy over the last years.
\begin{figure}[htp]
\vspace*{8cm}
\caption[graph]
{\small\it
\label{graph}
Diagrams describing a) $J/\psi$ radiative decay,
b) $J/\psi$ hadronic decays, c) $\gamma \gamma$ collisions.}
\end{figure}
However, up to now
the results obtained by the different experiments
are rather ambiguous.
This is mostly due to the complexity of the hadron spectrum, where
$q \bar q$  ground states overlap, in the same mass region, with radial
excitations so that, after some initial enthusiasm,
it now seems unlikely
that one single experiment could discover
gluonium states. The evidence for these non-$q \bar q$ mesons
can, in fact, only come from the comparison of light meson spectroscopy from
several dynamical sources, i.e. $J/\psi$ radiative and hadronic decays,
$\pi$ or K induced peripheral reactions, $p \bar p$ annihilation,
$\gamma \gamma$ collisions, central production etc.
\par
What characteristics are there which may help in disentangling glueballs
from quarkonium states?
\begin{itemize}
\item {a)}
Glueballs are flavour SU(3) singlets, so they have isospin zero and
are expected to couple, apart for phase space factors, equally to final states
of all flavours.
\item {b)}
Mesons are grouped into nonets with the same $J^{PC}$. The finding of
extra states having quantum numbers of an already completed nonet could be
a signal for having found an exotic state. However, the undefined situation of
radial excitations and multiquark states does not allow an easy classification
of newly discovered resonances.
\item {c)}
Glueballs and hybrids can have unusual production or decay characteristics.
\item {d)}
Glueballs and hybrids can have any $J^{PC}$ combination but some of
them (like $0^{+-}$, $0^{--}$, $1^{-+}$, ...) are not allowed for $q \bar q$
states. The finding of one of these states would be the best evidence for
the existence of gluonic mesons.
\item {e)}
Since gluons do not carry electric charge,
glueballs should not couple to $\gamma$'s,
A new parameter has been introduced to quantify this idea, the
"stickiness"~\cite{Chanow1}:
$$ S = {\Gamma(J/\psi \to \gamma X) \times PS(\gamma \gamma \to X)
\over PS(J/\psi \to \gamma X) \times \Gamma(\gamma \gamma \to X)}$$
This definition implies that S should be large for glueballs which
means that they should have large
branching ratios in "gluon rich" channels like radiative $J/\psi$ decay and
small $\gamma \gamma$ widths.
\end{itemize}
In conclusion, the strategy developed over the last years for finding
exotic mesons is based essentially on the following ideas:
\begin{itemize}
\item {i)}
Compare meson spectroscopy from different production mechanisms;
\item {ii)}
Make use of reactions which can "tag" the flavour content or the
quantum numbers of the produced resonance.
\end{itemize}

As an example of i) it is interesting to compare the $\eta \pi \pi$ mass
spectrum from radiative $J/\psi$ decay~\cite{jetapi} (fig.~\ref{etapi1}a))
with that from central hadronic collisions
in the reaction $p p \to p (\eta \pi \pi) p$~\cite{wetapi}
(fig.~\ref{etapi1}b)).
\begin{figure}[htp]
\vspace*{8cm}
\caption[etapi1]
{\small\it
\label{etapi1}
a) $\eta \pi \pi$ mass spectrum from radiative $J/\psi$ decay
(MarkIII), b) $\eta \pi \pi$ mass spectrum from central production
($\Omega$-WA76).}
\end{figure}
In radiative $J/\psi$ decay we observe a large production of
pseudoscalars, visible by the
strong enhancement in the
$\eta'$ region. In central production, on the other hand, axial vectors
(evidenced by the strong
$f_1(1285)$ signal) are seen to be enhanced with respect to pseudoscalars.

As an example of ii) and how $\gamma \gamma$ collisions selected in two
different
kinematic ranges are able to discriminate between different spin assignments
for the produced final states,
we show in fig.~\ref{etapi2} the $\eta \pi \pi$ mass spectrum from the
reaction $\gamma \gamma \to \eta \pi \pi$~\cite{getapi}.
\begin{figure}[htp]
\vspace*{8cm}
\caption[etapi2]
{\small\it
\label{etapi2}
a) $\eta \pi \pi$ effective mass from
$\gamma \gamma \to \eta \pi \pi$, b) $\eta \pi \pi$ mass distribution from
$\gamma \gamma^* \to \eta \pi \pi$. The data are from MarkII.}
\end{figure}
Fig. \ref{etapi2}a) shows the $\eta \pi \pi$
effective mass spectrum for "no tag" events, when the $Q^2$ of the reaction
sketched in fig.~\ref{graph}c) is so small that the two $\gamma$'s are real and
scattered electrons are not detectable because they are lost in the beam pipe.
Fig. \ref{etapi2}b), on the other hand, shows the same $\eta \pi \pi$ mass
spectrum
when one of the electrons is detected so that the $Q^2$ is relatively high
and one of the two $\gamma$ is not real. We observe the presence of the
$\eta'$ in both spectra but the spin-one resonance $f_1(1285)$ is visible
only in the
reaction $\gamma \gamma^* \to \eta \pi \pi$. The Yang-Landau theorem
states that two massless spin-one objects cannot combine to form a spin-one
object; thus, when a resonance is not seen in the fusion of two real
photons, but is observed when one of the photons is far from the mass shell, it
indicates that the resonance is probably spin-one.

\section*{\normalsize\bf 3. THE 0$^{-+}$, 0$^{++}$, 1$^{++}$ AND 2$^{++}$
MULTIPLETS}
\vskip 8pt

In the framework of the quark model, quark and anti-quark combine to multiplets
with well defined values of J$^{PC}$. States
with an angular momentum of L=1 between quark and anti-quark
in a spin-triplet state populate the
$^3$P$_0$ ($0^{++}$), $^3$P$_1$ ($1^{++}$) and $^3$P$_2$ ($2^{++}$)multiplets;
those with L=0 and the quark and anti-quark in a spin singlet state populate
the $^1$S$_0$ ($0^{-+}$) multiplet.
Even without assigning observed states to a given multiplet, the
existence of non-$\bar{\mbox{q}}$q states which lie in the same mass range
and have identical quantum numbers to those of these multiplets
can be established by the simple presence of more low-mass states
than predicted by the quark model.
Using the Godfrey-Isgur model as a guide, one can tentatively assign
the states listed in Table ~I with
those making up the 0$^{-+}$, 0$^{++}$, 1$^{++}$ and 2$^{++}$ multiplets,
although such an assignment is not necessarily unique. The
numbers in square brackets
are the Godfrey-Isgur predictions for each state.

\begin{table}[hbtp]\centering
\begin{tabular}{|c|c|c|c|} \hline
J$^{PC}$ & I=1 & I=0 & I=1/2 \\ \hline
0$^{-+}$ & $\pi$(140)  & $\eta$(547), $\eta$'(958) & K(494)   \\
         & [150]       & [520], [960]              & [470]     \\ \hline
0$^{++}$ & a$_0$(980)  & f$_0$(975), f$_0$(1400)   & K$^*_0$(1430) \\
         & [1090]      & [1090], [1360]            & [1240]    \\ \hline
1$^{++}$ & a$_1$(1260) & f$_1$(1285), f$_1$(1510)  & K$_{1A}$\footnotemark \\
         & [1240]      & [1240], [1480]            & [1380]    \\ \hline
2$^{++}$ & a$_2$(1320) & f$_2$(1270), f$_2$'(1525) & K$^*_2$(1430) \\
         & [1310]      & [1280], [1530]            & [1430]    \\ \hline
\end{tabular}
\end{table}
\footnotetext{
The K$_{1A}$ and the corresponding 1$^{+-}$ state K$_{1B}$ are nearly
45$^{\circ}$ mixed states of the K$_1$(1270) and K$_1$(1400).}

In this assignment, most J$^{++}$ states lie in the 1.3 GeV to 1.5 GeV mass
 region,
with a relatively small L$\cdot$S splitting. A notable exception are
the I=0 and I=1 0$^{++}$ states f$_0$(975) and a$_0$(980), which also
exhibit the largest mass differences to the Godfrey-Isgur
model predictions.

\section*{\normalsize\bf 4. 0$^{++}$ NON-$\bar{\mbox{q}}$q
 CANDIDATE STATES}
\vskip 8pt

Given the ambiguities in the assignments to the 0$^{++}$ nonet
and the discrepancies between the experimental states and the Godfrey-Isgur
predictions for this nonet, it is natural to consider alternative
assignments in which the isoscalar and isovector 0$^{++}$ states are identified
with some of the less-well determined resonances in the 1.3 -- 1.5 GeV region.
A consequence is that the f$_0$(975) might then be a weakly
bound $K\bar{K}$ system or of other non-$\bar{\mbox{q}}$q origin.
Information on this region
has been obtained by D. Morgan and M.R. Pennington~\cite{molec}. They have
combined data on $f_0(975)$ production in $J/\psi$ and $D_s$ decays with
information obtained from central production and elastic $\pi \pi$ and
$K \bar K$ processes.
Fig.~\ref{1GeVregion} shows the projection of their fits on the available
data from Mark III and DM2 assuming one (as for a $K\bar{K}$ molecule) or
two (as for a quark model state) poles.
In addition to
concluding that the $f_0(975)$ is most probably not a $K \bar K$ molecule
but that it has a conventional Breit-Wigner structure with a rather narrow
width ($\Gamma _0 \sim$ 52 MeV) and comparable couplings to $\pi \pi$ and
$K \bar K$, they argue the existence of an additional, very broad $f_0$(1000)
which would play the role of the lightest broad I=0 scalar.
Rather than reducing the ambiguities in the scalar sector however, this
increases the number of scalar states below 1800 MeV, since the existence
of a broad scalar structure around 1400 MeV seems experimentally well
established thanks to recent
observations of its decay to $\rho\rho$~\cite{f1400}, $\pi^0\pi^0$
and $\eta\eta$~\cite{CB:3pi0}.
A similar conclusion on the $f_0(975)$ is drawn from the study of centrally
produced $\pi \pi$ and $K \bar K$ final states but with slightly different
parameters ($\Gamma _0 = 72\pm 8$ MeV, $g_K/g_{\pi}=2.0 \pm 0.9$)~\cite{armpi}.
\begin{figure}[htp]
\vspace*{9cm}
\caption[1GeVregion]
{\small\it
\label{1GeVregion}
DM2 and MARK III data on J/$\psi \rightarrow \phi\pi^+\pi^-, \phi K^+ K^-$.
The fits with 1) one pole, 2) two poles are described in~\cite{molec}.
}
\end{figure}

On the other hand, new results from BES~\cite{cornell} detecting the f$_0$
recoiling against $\omega$ and against $\phi$ in J/$\psi$ decay
favor the molecular interpretation:
determining the relative decay ratios of
J/$\psi$ $\rightarrow$ $\phi$f$_0$(975)
to
J/$\psi$ $\rightarrow$ $\omega$f$_0$(975), they find a value of
2.2$^{+3.3}_{-0.8}$, consistent with the
molecular state expectation~\cite{seiden} of
BR(J/$\psi \rightarrow \phi f_0(975)) = 2
\cdot$ BR(J/$\psi \rightarrow \omega f_0(975))$.
They also determine a width
$\Gamma_{\pi\pi}$ = 36 $\pm$ 11 MeV for f$_0$(975), which is in good agreement
with the expected value~\cite{isgurf0} of $\sim$ 38 MeV for a
$\bar{\mbox{K}}$K molecular state.

The case for a molecular interpretation of the f$_0$(975) might be bolstered
by unambiguous evidence for further molecular states. One promising state
in this context is $\psi$(4040) which has been considered a
candidate for a D$^*\bar{\mbox{D$^*$}}$ molecule.
Since molecular states would be weakly bound, the decay pattern
should follow that of the quasi-free constituents, giving an easily
testable prediction for the decay branching ratios~\cite{barnes}.
Another experimental test is a determination of the
couplings of a$_0$ and f$_0$ to $\gamma\gamma$, which are predicted to be
small and equal in the molecular picture (with a large coupling to
$\bar{\mbox{K}}$K )~\cite{isgurf0}.

The existence of radial excitations of the  $\bar{\mbox{q}}$q states
increases the number of low mass states with a given J$^{PC}$.
The first radial excitation of the $^3$P$_0$ partner of the
0$^{++}$ ground state K$_0^*$(1430) has been observed in Kp scattering by LASS
at a mass of 1.95 GeV, and with a width of 200
MeV~\cite{Kradial}. The same experiment has also measured the
first radial excitation of the 2$^{++}$ K$_2^*$(1430) at a mass of
1.97 GeV and a width of 370 MeV in the process K$^-$p $\rightarrow$
$\overline{K^0}\pi^+\pi^-n$~\cite{Kradial}.
Similarly, the first radial excitation of the f$_2$(1270), possibly observed
in~\cite{Cason} with a mass of (1799 $\pm$ 15) MeV has a large width
of (280 $\pm$ 40) MeV. The general tendency of large widths and a mass
split of $\sim$ 500 MeV between the ground states and their first radial
excitation is also consistent with the predictions of the Godfrey-Isgur
model.
In particular, the first radial excitations in the 0$^{++}$ and
2$^{++}$ nonets, the 2$^3$P$_0$ and 2$^3$P$_2$ states, are predicted to lie
at 1780 MeV, resp. 1820 MeV.
\begin{figure}[htp]
\vspace*{9cm}
\caption[gammamgamma]
{\small\it
\label{gammagamma}
$\pi^+\pi^-$ (Mark II), $\pi^0\pi^0$ (Crystal Ball), K$_S$K$_S$ (CELLO/PLUTO)
and $K^+ K^-$ (ARGUS) invariant mass distributions from $\gamma\gamma$
collisions~\cite{gammagamma} )
}
\end{figure}

If the f$_0$(975) is not the lowest lying 0$^{++}$ state, its place in
the 0$^{++}$ nonet must be filled by some other ($\bar{\mbox{q}}$q) state.
The mass region 1.3 -- 1.6 GeV contains several more or less well established
candidate states; this mass window is however somewhat higher than
the prediction from the model of Godfrey and Isgur.

There are indications from LASS of an S-wave structure at 1.53 GeV
in K$^-$p $\rightarrow K_S^0K_S^0\Lambda$~\cite{aston1988},
which would be naturally interpreted as the $^3$P$_0$ ground state
partner of the f$_2$'(1525).
Indications of the corresponding iso-vector state
have been seen in its neutral mode in $\eta\pi^0$ by GAMS~\cite{gams}, as
well as in its charged mode in $\eta\pi^-$ by Benkei~\cite{aoyagi} at
about 1.3 GeV in $\pi$p $\rightarrow \pi\eta$n (Fig.~\ref{benkei}).
It should be noted that in spite of larger statistics, VES sees no activity
around 1.3 GeV in the S-wave in a partial
wave analysis of the $\eta\pi^-$ system produced in $\pi$N scattering
at 37 GeV/c~\cite{borisov}.
\begin{figure}[htp]
\vspace*{8cm}
\caption[benkei]
{\small\it
\label{benkei}
a) Results from the $\eta\pi^0$ partial wave analysis
from the reaction $\pi^-$p $\rightarrow \eta\pi^0$n (GAMS).\\
b) Results from the $\eta\pi^-$ partial wave analysis
from the reaction $\pi^-$p $\rightarrow \eta\pi^-$p~\cite{aoyagi}.\\
c) Results from the K$^0_S$K$^0_S$  partial wave analysis
from the reaction K$^-$p $\rightarrow$ K$^0_S$K$^0_S\Lambda$ (LASS).
}
\end{figure}
The fact that these states would be mass-degenerate with
f$_2$(1530) and a$_2$(1320), together with the relatively low statistical
significance of the signals, gives rise to the worry of feedthrough from
the partial wave analyses;
confirmation is needed for both states. An I=1 O$^{++}$ state
a$_0$(1430 $\ldots$ 1480) with a width of 230 $\ldots$ 270 MeV has been
observed by the Crystal Barrel collaboration~\cite{como} in the
process $\bar{\mbox{p}}$p $\rightarrow \eta\pi^0\pi^0$; further
study is needed, but
it would be important to ascertain whether this state could be the
first radial excitation or the $^3$P$_0$ ground state itself. In both
cases, this state could set the mass region of the 0$^{++}$ nonet.

Recently, a very broad (4$\pi$)$^0$ enhancement has been seen in
nucleon-antinucleon annihilation into five pions by two
groups~\cite{f1400}.
In both cases, a dominant decay mode of this
I=0 0$^{++}$ object is $\rho\rho$, although it is also
observed in $\pi^0\pi^0$, $\eta\eta$~\cite{CB:3pi0} and
$\sigma\sigma$.
The mass (1374 $\pm$ 38 MeV) and width ( 375 $\pm$ 61 MeV)
of this state are compatible with the f$_0$(1400), but its decay modes --
which are those expected for a ($u\bar{u}+d\bar{d}$)
state in the same nonet as K$_0^*$(1430) -- are not;
here too, further study is needed.

Two states with J$^{PC}$ = 0$^{++}$ lie in the 1500 - 1600 MeV mass region.
The state G(1590), first observed by the GAMS experiment
in the process $\pi^-$p $\rightarrow \eta\eta$n and
$\eta\eta'$n~\cite{gamsG1590}, has
unusual decay properties, in that its decay rate
into $\eta\eta'$ is three times larger than its $\eta\eta$ decay,
and in that it has not been seen in K$\bar{\mbox{K}}$.
An upper limit for BR(G(1590) $\rightarrow \pi^0\pi^0$) of less than
0.3 $\times$ BR(G(1590) $\rightarrow \eta\eta$) is also
found~\cite{gamsG1590}.

The Crystal Barrel group has observed a 0$^{++}$ state~\cite{como}
with a mass of 1520 $\pm$ 45 MeV and a width of 148 $\pm$ 25 MeV
decaying into $\pi^0\pi^0$ and $\eta\eta$ in
$\bar{\mbox{p}}$p  annihilation at rest (Fig.~\ref{3pi0DP}).
At the same time, they give an upper limit for the ratio of decay ratios
of a state around 1550 MeV into $\eta\eta'$ and into $\eta\eta$ of
less than 0.2~\cite{como}.
\begin{figure}[htp]
\vspace*{12cm}
\caption[3pi0DP]
{\small\it
\label{3pi0DP}
a) Dalitz plot for 3$\pi^0$ events from Crystal Barrel.
b) $\pi^0\pi^0$ invariant mass distribution (the solid line corresponds
to a preliminary fit containing the f$_0$(1515)).
The scalar amplitude f$_0$(1515) corresponds to
the narrow bands that cross the Dalitz plot at $\sim$ 2.3 GeV$^2$.
c) $\eta \eta$ effective mass from $\bar p p \to \pi^0 \eta \eta$
}
\end{figure}
No indications for a scalar state in the same mass region of $\sim$ 1550 MeV
come from radiative J/$\psi$ decay~\cite{chen},
nor central production~\cite{palano}, while a weak signal around
1.5 GeV is seen in the S-wave in Kp scattering~\cite{aston1988}, which would
speak against a glueball interpretation. On the other hand, QCD sum rules
predict a suppression by one order of magnitude for scalar glueballs relative
to tensor glueballs~\cite{scalar2tensor}, so that given the available
statistics,
non-observation in J/$\psi$ radiative decays would not be surprising.

Assuming the above assignment of f$_0$(1400), f$_0$(1520) could then only be
assigned to the ninth member of the 0$^{++}$ nonet, mostly $s\bar{s}$, which
would however contradict the observed strong coupling to $\pi\pi$, as well
as the weakness of this state in Kp scattering.
If the Crystal Barrel f$_0$(1520) is identified with the GAMS f$_0$(1590),
then the two experiments are in contradiction with respect to the $\eta\eta'$
and $\pi^0\pi^0$ decay modes.
If on the other hand, the GAMS f$_0$(1590) decay to $\eta\eta'$
really is dominant, then there is an excess of 0$^{++}$ states around 1500 MeV.
The conclusion that one may be a non-$\bar{\mbox{q}}$q state, or that the
f$_0$(1520) and f$_0$(1590) are mixed states of a glueball and the ninth
member of the $q\bar{q}$ nonet is tempting.
In this respect, the absence of a signal for either state in $\pi^0\pi^0$
in $\gamma\gamma$ collisions~\cite{gammagamma} is significant.
A second possibility would be that of identifying f$_0$(1520) with the
first radial excitation, i.e. the 2 $^3$P$_0$ state; however, its relatively
narrow width of 148 $\pm$ 25 MeV speaks against this possibility.
While the search for further decay modes of the f$_0$(1500) and/or f$_0$(1590)
will help in clarifying their nature, a high statistics search for either
state in J/$\psi$ radiative decay, as well as a clarification of the nature
of the f$_0$(975), would clearly be of great importance to
establish their glue content.
\begin{figure}[htp]
\vspace*{8cm}
\caption[g1590]
{\small\it
\label{g1590}
Evidence for the G(1590) from GAMS.
a) $\eta\eta$ invariant mass distribution.\\
b) S-wave contribution from a partial wave analysis of the $\eta\eta$
invariant mass distribution.\\
c) $\eta\eta'$ invariant mass distribution. The dashed line is the phase space
normalized to the number of events in the measured mass interval,
the full line a Breit-Wigner fit~\cite{gamsG1590}.
}
\end{figure}
\section*{\normalsize\bf 5. 2$^{++}$ NON-$\bar{\mbox{q}}$q
 CANDIDATE STATES}
\vskip 8pt
The state $\theta$(1720) has been of considerable interest since its
discovery by Crystal Ball through the process
J/$\psi \rightarrow \gamma\theta, \theta \rightarrow \eta\eta$,
and its subsequent confirmation by MARK III and DM2
(in radiative J/$\psi$ decays), as well as by fixed target experiments
($\Omega$-WA76). Its J$^{PC}$, originally determined to be 2$^{++}$
was later revised to 0$^{++}$ by MARK III~\cite{chen}.
Recent J$^{PC}$ determinations in central production and
J/$\psi$ radiative decay~\cite{wa76,cornell}
however again favor a 2$^{++}$ assignment.
Fig.~\ref{wa76} shows the signal and decay angular distributions of
 $\theta$(1720)
in central production; both f$_2$(1525) and $\theta$(1720) are well
described by a 2$^+$ angular distribution. An analysis by BES
of the processes J/$\psi \rightarrow \omega K^+K^-$ and
J/$\psi \rightarrow \gamma \pi^0\pi^0$ gives a dominance of 2$^{++}$
in this mass region, but allows for a 0$^{++}$ component on the high mass
side (around 1750 MeV) in the spin analysis of the reaction
$J/\psi \to \gamma K^+ K^-$. Assuming the 2$^{++}$ assignment is correct,
then $\theta$ cannot be the $s\bar{s}$ member of the 2$^{++}$ nonet
(a role fulfilled by f$_2$'(1525)), but is too light to be a radial excitation.
\begin{figure}[htp]
\vspace*{9cm}
\caption[wa76]
{\small\it
\label{wa76}
K$\bar{\mbox{K}}$ invariant mass distributions from a) WA76, b) BES.
Angular distributions for c) f$_2$'(1525) and d) $\theta$(1720).
K$\bar{\mbox{K}}$ invariant mass distributions recoiling against
e) $\omega$ and f) $\phi$ from MARK III.
}
\end{figure}
Several other properties also make the $\theta$(1720) a rather unique state.
Although the $\theta$ decays predominantly to $K\bar{K}$~\cite{PDG}
(and to a lesser degree to $\pi\pi$ and $\eta\eta$), it is
not produced in Kp scattering~\cite{aston1988}, and is produced at a much
larger
rate in radiative J/$\psi$ decay than the
$s\bar{s}$ f'$_2$(1525)~\cite{koepke}.
This is also seen in central production where the $t$
distribution for f$_2$'(1525) and $\theta$(1720) are quite
different~\cite{wa76}.

The comparison of J/$\psi$ decays into $\omega$+X and $\phi$+X allows a
determination of the quark content of the resonance X. In the approximation
of ideal mixing, a state X recoiling against an $\omega$ will consist of
$u$ and $d$ quarks, while a state recoiling against a $\phi$ contains strange
quarks. A clear signal for $\theta$ is seen in K$^+$K$^-$ recoiling against
$\omega$. However, the $K^+ K^-$ spectrum recoiling against a $\phi$ shows
the expected presence of f$_2$'(15252) but it is not clear if the shoulder
visible in the 1.65 region can be attributed to the presence of the
$\theta(1720)$.




Although many states are expected in the 2 GeV region, several
structures seen around 2.2 GeV in radiative J/$\psi$ decays
are remarkable by their unexpectedly narrow width (smaller than the
experimental resolutions of $\sim$ 100 MeV); with 0$^{++}$
glueball candidates around 1.5 GeV setting the mass scale, a 2$^{++}$
glueball is expected in this mass region.
A narrow state at 2.23 GeV, first seen by MARK III in radiative J/$\psi$
decay to K$^+$K$^-$ and K$_S$K$_S$~\cite{zeta2230},
has been recently confirmed by BES~\cite{cornell},
which has also observed this
state in the process J/$\psi \rightarrow \gamma\eta\eta$.
A spin-parity analysis by MARK III gives J$\ge$2, while J$^{PC}$ must lie
in the series (even)$^{++}$ due to the observation in K$_S$K$_S$.
A weak structure at 2.22 GeV is also seen in $\pi^-$p scattering
in $\eta\eta'$.
A lower limit J $\ge$ 2 is obtained from anisotropic angular distributions.
In view of the strange quark content of $\eta$ and $\eta$', it is tempting
to identify this structure with the $\xi$(2230) above.
A spin-parity analysis of the signal seen
at BES is in progress, but may be limited by statistics as well as
detector acceptance. This analysis might be able to differentiate between
spin 0 and 2, but will most likely not be able to test the
suggestion that $\xi$ is the 4$^{++}$ $s\bar{s}$ partner of the
f$_4$(2030)~\cite{gkp}, as
is suggested by an analysis of LASS data which finds a
a 4$^{++}$ resonance at 2.209 GeV in the process
K$^-$p $\rightarrow$ K$_S$K$_S$$\Lambda$~\cite{aston1988}.
\begin{figure}[htp]
\vspace*{9.5cm}
\caption[jetset]
{\small\it
\label{jetset}
a) K$_S$K$_S$ and $\eta\eta$ invariant mass from radiative J/$\psi$ decay from
 BES.
b) $\phi\phi$ invariant mass from radiative J/$\psi$ decay from MARK III
(efficiency-corrected spectrum with fits to Breit-Wigner resonances).
c) Cross section for the process $\bar{\mbox{p}}$p $\rightarrow$ K$_S$K$_S$.
}
\end{figure}
A possibly different state at the same mass and
with a comparable width is seen in
radiative J/$\psi$ decay to $\phi\phi$ by MARK III
and DM2~\cite{phiphiMII}.
It is interesting to note that while spin-parity analyses of both experiments
suggest a pseudoscalar assignment for this state, spin-parity analyses
of a $\phi\phi$ resonance in the same mass region in
hadronic production~\cite{wa67}
are consistent with J$^P$ = 2$^+$.
In view of $\omega-\phi$ mixing, it is natural to complement the study of
this state by a search in the processes
J/$\psi \rightarrow \omega\phi$  and
J/$\psi \rightarrow \omega\omega$. The former in particular would be
interesting
as it could be an indication for a ($u\bar{u}+d\bar{d}$)g hybrid;
such a hybrid with J$^{PC}$ = 2$^{++}$ is predicted at 2.32
GeV~\cite{chansharpe}.

It is all the more intriguing that no evidence for
either state has been found in a $\bar{\mbox{p}}$p formation experiment.
In none of the reactions $\bar{\mbox{p}}$p $\rightarrow$ K$^+$K$^-$,
$\bar{\mbox{p}}$p $\rightarrow$ K$_S$K$_S$ or
$\bar{\mbox{p}}$p $\rightarrow$ $\phi\phi$ is there an indication of
resonant behaviour in the 2.2 GeV region~\cite{jetset}. Fig.~\ref{jetset}
shows the measured cross-section for $\bar{\mbox{p}}$p $\rightarrow$
K$_S$K$_S$.

\section*{\normalsize\bf 6. PSEUDOSCALARS}
\vskip 8pt
After the initial discovery of the $\iota/\eta(1440)$ signal in $J/\psi$
radiative decay several experiments looked to the 1.4 GeV mass region
in a variety of different reactions, from $p \bar p$ annihilations to
$\pi$ or K induced reactions, from central production to $\gamma \gamma$
collisions. However, the striking complication appeared
that in the $E/\iota$ mass region the number of states
which contributed to the enhancement observed in the mass spectrum
changed from one experiment to the other.
This confusing experimental situation has led,
in the last ten years, to an intense
phenomenological debate (the $E/\iota$ puzzle) on the possibility that
one or more of these
states are non-$q \bar q$ mesons such as glueballs, hybrids or
multiquark states~\cite{palchu}.
Finally, a recent partial wave analysis of the $\iota$ region
performed by
the MarkIII group (see fig.~\ref{markpwa})
on the $K \bar K \pi$ and $\eta \pi \pi$ final states
from $J/\psi$ radiative decay,
has shed a new light on this puzzle~\cite{baipwa}.
\begin{figure}[htp]
\vspace*{8cm}
\caption[markpwa]
{\small\it
\label{markpwa}
a) Partial wave analysis of $J/\psi \to K \bar K \pi$;
b) $J^{PC}=0^{-+}$ intensity from the PWA of
$J/\psi \to \gamma \eta \pi \pi$ (MarkIII).}
\end{figure}
This analysis interprets the $\iota$ signal as due to three different
resonances
\begin{itemize}
\item {i)}
$\eta(1420)$ with $J^{PC}=0^{-+}$ and decay via $a_0(980)\pi$.
Very likely it has also a substantial $\rho^0 \gamma$ decay
mode~\cite{coff,rhog}
(see fig.~\ref{rhogam}a)).
This state may be the same as the one
observed in $p \bar p$ at rest and in $\pi^-$
induced reactions. However, it is not observed in $\gamma \gamma$ collisions
nor in central production.
\item {ii)}
$\eta(1490)$ with $J^{PC}=0^{-+}$ and a decay via $K^* \bar K$
\item {ii)}
$f_1(1440)$ with $J^{PC}=1^{++}$ and a decay via $K^* \bar K$
\end{itemize}
\begin{figure}[htp]
\vspace*{8cm}
\caption[rhogam]
{\small\it
\label{rhogam}
a) $\rho^0 \gamma$ mass spectrum from $J/\psi$ radiative
decay (MarkIII);b) $\rho^0 \gamma$ mass spectrum from central production
($\Omega$-WA76).
}
\end{figure}
In conclusion, one or more pseudoscalars are present
in the 1.4 GeV mass region,
and up to now it is not clear whether they are
radial excitations, hybrids or glueballs.
\vskip 21pt
\section*{\normalsize\bf 7. AXIAL VECTORS}
\vskip 8pt
One of the most interesting mesons from the point of view of the
possible existence of non-$q\bar q$ states is the $J^{PC}=1^{++}$
$E/f_1(1420)$ meson.
\par
The best evidence for $E/f_1(1420)$ comes from the $\Omega-WA76$ experiment
which studied the reaction
$p p \to p (K \bar K \pi) p$~\cite{ewa76}.
The same experiment also studied the centrally produced $4 \pi$~\cite{pi4wa76}
(fig.~\ref{ex1450}),
$\eta \pi \pi$~\cite{wetapi} (fig.~\ref{etapi1}b)) and
$\rho^0 \gamma$~\cite{rhog}
(fig.~\ref{rhogam}b)) final states.
While the
presence of the axial meson $f_1(1285)$ was observed in all these spectra,
the $f_1(1420)$ was found to decay only to $K^* \bar K$.
The classification of the $E/f_1(1420)$ in the quark model is still unclear,
its quantum
numbers are sometimes subjected to criticism and its interpretation
as a normal hadronic resonance is not without problems.
It was considered,
until recently, to be the $s \bar s$ member of the axial nonet.
However, this hypothesis is in contradiction to several experimental
results, namely:
\begin {itemize}
\item {i)} It is not produced in $K^-$ induced reactions, where
an $s \bar s$ state should prominently appear. On the other hand
a different axial resonance, the $f_1(1520)$, has been discovered
in these reactions~\cite{dprime}, which has the expected properties for being
the $s \bar s$ member of the axial meson nonet (see fig.~\ref{fewa76}b).
\item {ii)} The pattern observed in hadronic $J/\psi$ decay
($J/\psi \to \omega E$ seen, $J/\psi \to \phi E$ not seen)~\cite{becker}
(see fig.~\ref{eall}a,c)
is inconsistent with a mainly $s \bar s$ composition of the
$E/f_1(1420)$ meson. The same conclusion is obtained from the
observed rates for production of this resonance in $\gamma \gamma^*$
collisions (see fig.~\ref{eall}):
$\Gamma_E$ is too large for a mainly strange meson~\cite{caldwell}.
\end {itemize}
\begin{figure}[htp]
\vspace*{8cm}
\caption[fewa76]
{\small\it
\label{fewa76}
a) $K \bar K \pi$ mass spectrum centrally produced in pp
interactions ($\Omega$-WA76);
b) $K \bar K \pi$ mass spectrum from an incident
$K^-$ beam (LASS).
}
\end{figure}
These arguments lead to two possibilities: either the $E/f_1(1420)$
belongs to the axial nonet with a mixing angle far from the ideal
one leaving the $f_1(1520)$ as an extra state or, more reasonably,
the $E/f_1(1420)$ is an extra resonance which does not fit
into the quark model.
\begin{figure}[htp]
\vspace*{11cm}
\caption[eall]
{\small\it
\label{eall}
Comparison between the $K \bar K \pi$ mass spectra
from:
a)$J/\psi \to \gamma K \bar K \pi$,
b)$J/\psi \to \omega K \bar K \pi$,
c)$J/\psi \to \phi K \bar K \pi$ (MARKIII);
d) $\gamma \gamma \to K \bar K \pi$,
e) $\gamma \gamma^* \to K \bar K \pi$ (TPC/$2\gamma$)
}
\end{figure}
In the latter case it is interesting to understand what
it really is: a hybrid meson (see fig.~\ref{ishida}~\cite{ishida}),
a $K^* \bar K$
or a multiquark state~\cite{caldwell,longacre}?
\begin{figure}[htp]
\vspace*{4cm}
\caption[ishida]
{\small\it
\label{ishida}
Possible decay of a hybrid meson to $K^* \bar K$
}
\end{figure}
\section*{\normalsize\bf 8. THE SEARCH FOR THE $J^{PC}=1^{-+}$ EXOTIC STATES}
\vskip 8pt
The discovery of an exotic $1^{-+}$ combination, which is impossible to form
with quarks only, would give a strong push to gluonium spectroscopy.
For this reason great interest was provoked by the claim of the GAMS
experiment of having found one of these states in the $\eta \pi^0$ mass
distribution from incident $\pi^-$ beams~\cite{rhogams}. The $\eta \pi^0$
mass spectrum from this experiment
is dominated by a large $a_2(1310)$ resonance.
However, a partial wave
analysis of the $\eta \pi^0$ mass spectrum revealed the existence, below
the large tensor wave, of a smaller but significant spin 1 wave interpreted
as the evidence of a $1^{-+}$ resonance ($\rho(1406)$) having a mass of 1406
MeV
and $\Gamma$=180 MeV (fig.~\ref{exot}).
\begin{figure}[htp]
\vspace*{8cm}
\caption[exot]
{\small\it
\label{exot}
a) Results from the $\eta \pi^0$ PWA: spin-1 wave (GAMS);
b) $\eta \pi^-$ mass spectrum from the reaction
$\pi^- p \to \eta \pi^- p$ at 6.3 GeV/c (KEK).
}
\end{figure}
The presence of a $1^{-+}$ wave in the 1.4 mass region has been confirmed
at KEK~\cite{aoyagi} in the study of the $\eta \pi^-$ final state (fig. 10c),
but with somewhat different parameters.
However, there are some problems with the analysis method which may yield
ambiguous results showing the need of confirmation in different processes.
\par
Other experiments have investigated the $\eta \pi$ spectrum in a search for
this exotic resonance. In particular the VES experiment, at IHEP
has collected large statistics on
the reactions $\pi^-N \to (\eta \pi^-) N$
and $\pi^-N \to (\eta' \pi^-) N$
finding no evidence for exotic resonance production in
the $\eta \pi^-$ and $\eta' \pi^-$ final states (fig. 11a).
The same result has been obtained
by the Crystal Barrel experiment at Lear which studied the $\eta \pi^0$
system in the reaction $\bar p p \to \pi^0 (\eta \pi^0)$
(fig.~\ref{exot1}b)~\cite{como}.
\par
The exotic isospin zero hybrid ($\omega(1^{-+})$)
is expected by several models
to be in the 1.3-1.6 GeV mass region~\cite{gutsche} and to have an important
$a_1(1260) \pi$ decay mode.
Fig.~\ref{ex1450} shows the $2\pi^+ 2\pi^-$ mass distribution
from $\Omega-WA76$ experiment in the reaction
$p p \to p (2\pi^+ 2\pi^-) p$~\cite{pi4wa76}.
\begin{figure}[htp]
\vspace*{8cm}
\caption[exot1]
{\small\it
\label{exot1}
a) Results from VES: $1^{-+}$ waves;
b) $\eta \pi^0$ effective mass  from
$\bar p p \to \eta \pi^0 \pi^0$ (Crystal Barrel)}
\end{figure}
Along with $f_1(1285)$ and a broad structure peaking at 1900 MeV, a new
resonance has been discovered: X(1449) having a mass of 1449 MeV
and a width of 80 MeV. It decays to $\rho^0 \pi \pi$ and a spin analysis
favours $J^{PC}=0^{++}$, $1^{-+}$ and $2^{++}$. However the size of the
background under the peak makes very difficult a reliable spin-parity analysis.
Evidence for a spin one resonance in this mass region has also been
reported in $\gamma \gamma$ collisions~\cite{bauer}.
\begin{figure}[htp]
\vspace*{8cm}
\caption[ex1450]
{\small\it
\label{ex1450}
$2\pi^+ 2\pi^-$ effective mass distribution from
$\Omega$-WA76 experiment.
}
\end{figure}
\section*{\normalsize\bf 9. EXOTIC RESONANCES WHICH DECAY TO VECTOR-VECTOR}
\vskip 8pt
The study of associated $\phi \phi$ production in $\pi^-$p interactions
was one of the starting points of gluonium spectroscopy.
The large and unexpected $\phi \phi$ cross section in $\pi^-$ p
interactions has been interpreted as due to the production of
three $J^{PC}=2^{++}$ glueball states in the 2.0 - 2.5 GeV region~\cite{etkin}.
Evidence for resonant structures in the 1.6 and 2.0 GeV regions
has been recently reported in the
$\omega \omega$ system produced by incident $\pi^-$ beams~\cite{omega2}.
\par
One of the most striking effects found in two photon physics is the
large cross section for $\gamma \gamma \to \rho^0 \rho^0$
below threshold~\cite{rhorho}. The much lower $\rho^+ \rho^-$ cross section
(fig.~\ref{vec1})
rules out a single resonance interpretation.
To explain this effect in a resonance interpretation requires the
introduction of an exotic I=2 state interfering with another I=0
state, both states being logical candidates for four-quark resonances.
Experimental spin parity analyses of the $\rho \rho$ enhancement are
controversial; it may be $0^+$ and/or $2^+$.
\begin{figure}[htp]
\vspace*{8cm}
\caption[vec1]
{\small\it
\label{vec1}
$\rho^0 \rho^0$ and $\rho^+ \rho^-$ cross sections
in $\gamma \gamma$ collisions
}
\end{figure}
As regards the $J/\psi$ radiative decays, the
$\phi \phi$, $\omega \omega$, $K^{*0} \bar K^{*0}$ and
$\rho \rho$ final states are dominated by $J^P=0^-$ contributions.
These spectra show marked threshold enhancements,
whereas the $\rho \rho$ final state
shows the presence of still unexplained resonant structures~\cite{rhorhom}
(see fig.~\ref{vec2}).
\begin{figure}[htp]
\vspace*{9.5cm}
\caption[vec2]
{\small\it
\label{vec2}
a) $\phi \phi$, b) $K^* \bar K^*$,
c) $\omega \omega$ and d),e)$\rho^0 \rho^0$ mass distributions
from radiative $J/\psi$ decay (MarkIII and DM2).
}
\end{figure}
\section*{\normalsize\bf 10. PROSPECTS FOR A TAU-CHARM FACTORY}
\vskip 8pt
The unique identification of a glueball or an hybrid state would certainly
considered as a fundamental discovery. However, up to now, the situation
is rather ambiguous mostly due to the lack of high statistics and high
precision data. Simplified analysis of data coming from low acceptance
detectors have created much confusion.
In this context a Tau-Charm Factory has a good chance to definitively solve the
problem of the existence of gluonic mesons~\cite{slac}.
Several
different tools would be available at a Tau-Charm Factory:
\begin {itemize}
\item {1)}
Radiative $J/\psi$ decays;
\item {2)}
Hadronic $J/\psi$ decays;
\item {3)}
$\eta_c$ and $\chi$ decays;
\item {4)}
$\gamma \gamma$ collisions.
\end {itemize}

There are several advantages in using $J/\psi$ to study light meson
spectroscopy. These are the following:

\begin {itemize}
\item {a)}
$J/\psi$ has well defined initial quantum numbers and is produced with
almost no background in $e^+ e^-$. This allows one to perform
reliable spin parity
analyses with a small number of amplitudes.
\item {b)}
At a Tau-Charm Factory it could be possible to easily obtain very large
statistics.
\item {c)} Its mass is ideal for exploring masses up to 2.5 GeV. Its decay
patterns involve gluons (so that glueballs could be formed) and mixtures of
quarks and gluons (for searching for hybrids).
\item {d)}
By comparing rates for $J/\psi \to \gamma + M_1$ to those for
$J/\psi \to M_1 + M_2$, $\eta_c \to M_1 + M_2$, $\chi_{c0,1,2} \to M_1 + M_2$
one can determine the spin and the quark/gluon content of a given resonance.
\end {itemize}

The actual number of $J/\psi$ decays collected up to now by
several experiments are summarized
in fig.~\ref{jpsi} and they do not exceed $10^7$.
At a Tau-Charm Factory this number
could easily grow to $10^9$ and simultaneously it could be possible to obtain
of the order of $10^7$ $\eta_c$ or $\chi$ decays through the chains:
$$J/\psi \to \gamma \eta_c$$
$$\psi' \to \gamma \chi_{c0,1,2}$$
The radiative $\psi'$ decays to $\chi$'s have quite large branching rations,
between 8 and 9 $\%$.
\begin{figure}[htp]
\vspace*{7cm}
\caption[jpsi]
{\small\it
\label{jpsi}
Number of collected
$J/\psi$ decays from the different experiments ($\times 10^6$).
}
\end{figure}
\par
Two photon physics is an important laboratory for studying light meson
spectroscopy. Glueballs should not be produced but hybrids and four-quark
resonances are accessible. Therefore, by comparing results from $J/\psi$
decays to those coming from photon-photon physics, it is possible to obtain
further information on the properties of exotic candidates.
\par
At a Tau-Charm Factory, the presence of an electromagnetic calorimeter at small
angles allows the detection of single and double tag events. This is a unique
possibility among the existing machines, even if the limited center of mass
energy allows the detection of resonances only in the low mass region, up to
2 GeV.

\section*{\normalsize\bf 11. CONCLUSIONS}
\vskip 8pt
It is now 13 years since the discovery of the glueball candidate $\iota(1440)$
in radiative $J/\psi$ decay.
Due to a large amount of experiments performed at a large
variety of fixed target and collider experiments,
this frontier of physics has advanced
considerably in the last years.
At present there are some "solid" candidates, but
the unambiguous identification of glueballs or
hybrids is still missing.

In all the $J^{PC}$ = 0$^{++}$, 0$^{-+}$, 1$^{++}$ and 2$^{++}$ sectors,
the number of observed states in the 1 -- 2 GeV mass region
is possibly larger than predicted by the quark model. The extra states
-- if their J$^{PC}$ are confirmed -- are
good candidates for non-$\bar{\mbox{q}}$q states,
but their nature will not be elucidated without a concerted search
in a large number of production mechanisms. The validation of any state as
glueball is strongly dependent on its observation in the decay of
J/$\psi$'s copiously produced at a $\tau$-charm factory.
Very large data sets (for reliable partial wave analyses) and
a detector covering as large a solid angle as possible (to minimize
distortions of decay angular distributions due to acceptance) are requisites
for observation and accurate determination of the
J$^{PC}$ of all candidate states, and for an unambiguous
identification of glueballs.
\par
The next decade should possibly solve this QCD low energy puzzle by using
high quality and high statistics data. A Tau-Charm factory is probably one of
the best places where this type of research can be performed.
\vskip 21pt

\section*{\normalsize\bf Acknowledgments}
We thankfully acknowledge discussions with
T.~Barnes,
D.~Bauer,
F.~Close,
A.~Falvard,
M.~Feindt
and R.~Landua.


\vskip 30pt
\begin{thebibliography}{99}

\bibitem{Isgur}
S.~Godfrey and N.~Isgur, Phys. Rev. D32 (1985) 189.


\bibitem{UKQCD}
G.S.~Bali et al. (UKQCD Collaboration), Phys. Lett. B 309 (1993) 378.

\bibitem{glueballpredictions}
see for example F.~Close, Rep. Prog. Phys. 51 (1988) 833 for a general review.


\bibitem{chansharpe}
M.S.~Chanowitz and S.R.~Sharpe, Phys. Lett. 132 B (1983) 413.

\bibitem{iota1}
D.L. Scharre et al., Phys. Lett. 97B (1980) 329.

\bibitem{iota2}
C. Edwards et al., Phys. Rev. Lett. 49 (1982) 259.

\bibitem{theta}
C. Edwards et al., Phys. Rev. Lett. 48 (1982) 458.

\bibitem{Chanow1}
M. Chanowitz, Proc. of VI Int. Workshop on Photon-Photon Collisions,
Tahoe, ed. R. Lander (World Scientific, 1984).

\bibitem{jetapi}
J. Becker et al., SLAC-PUB-4246 (1987).

\bibitem{wetapi}
T.A. Armstrong et al., Z. Phys. C52 (1991) 389.

\bibitem {getapi}
G. Gidal et al., Phys. Rev. Lett. 59 (1987) 2012.

\bibitem{Yang}
L.F. Landau, Dok. Akad. Nauk USSR 60 (1948) 207;
C.N. Yang, Phys. Rev. 77 (1950) 242.


\bibitem{molec}
Morgan and Pennington, Phys. Lett. B258 (1991) 444.\\
Morgan and Pennington, Phys. Rev. D48 (1993) 1185.

\bibitem{f1400}
C.~Amsler et al., Phys. Lett. B322 (1994) 431;\\
Obelix Collaboration, Proc. of Hadron '93, Como, Italy, June 1993.\\
M. Gaspero, Nucl. Phys. A562 (1993) 407.

\bibitem{CB:3pi0}
C. Amsler et al., Phys. Lett. B (1994) in press.

\bibitem{armpi}
T.A. Armstrong et al., Z. Phys. C51 (1991) 351.

\bibitem{cornell}
Zheng Zhipeng, XVI Inter. Symp. on Lepton-Photon Interactions, Cornell, 1993.

\bibitem{seiden}
A.~Seiden, VII Inter. Workshop on Photon-Photon Collisions, Paris, 1986.

\bibitem{isgurf0}
N.~Isgur, Phys. Rev. D 41 (1990) 2236.

\bibitem{barnes}
T.~Barnes, K.~Dooley and E.S.~Swanson, Phys. Lett. B275 (1992) 478.

\bibitem{Kradial}
D.~Aston et al., Nucl. Phys. B296 (1988) 493.\\
D.~Aston et al., Proc. of the Inter. Conf. on Hadron Spectroscopy,
Maryland, 1991.

\bibitem{Cason}
N.M.~Cason et al., Phys. Rev. D28 (1983) 1586.
\bibitem{Kada}
E.~Kada, P.~Kessler, and J.~Parisi, PRD 39 (1989) 2657.

\bibitem{gammagamma}
M.~Feindt and J.~Harjes, Nucl. Phys. B (Proc. Suppl.) 21 (1991) 61,
and references therein.

\bibitem{aston1988}
D.~Aston et al., Phys. Lett. 215B (1988) 199.

\bibitem{gams}
M.~Boutemer and M.~Poulet, Proc. 3$^{rd}$ Inter. Conf. on Hadron Spectroscopy,
Ajaccio, 1989.

\bibitem{aoyagi}
H.~Aoyagi et al., PLB 314 (1993) 246.

\bibitem{borisov}
G.V.~Borisov et al., Proc. of Hadron '93, Como, Italy, June 1993.

\bibitem{como}
M.~Doser at al., Proc. of Hadron '93, Como, Italy, June 1993.

\bibitem{gamsG1590}
F.~Binon et al., Nuovo Cim. 78A (1983) 313;
F.~Binon et al., Nuovo Cim. 80A (1983) 363;
D.~Alde et al., Phys. Lett. B201 (1988) 160.
D.~Alde et al., Nucl. Phys. B269 91986) 485.


\bibitem{chen}
L.-P.~Chen et al., Nucl. Phys. B (Proc.Suppl.) 21 (1991) 80.

\bibitem{palano}
A.~Palano et al., Nucl. Phys. B (Proc.Suppl.) 21 (1991) 49.

\bibitem{scalar2tensor}
V.A.~Novikov, M.A.~Shifman, A.I.~Vainshtein and V.I.~Zakharov, Nucl. Phys. B165
(1980) 55,67; B191 (1981) 301.

\bibitem{wa76}
T.A.~Armstrong et al., Phys. Lett. B 227 (1989) 186.

\bibitem{PDG}
Particle Data Group, Phys. Rev. D 45 (1992).

\bibitem{koepke}
L.~K\"opke, p.692, Proc. XXXIII Int'l. Conf. on H.E.P. (World Scientific,
Singapore, 1986, ed. S.~Loken)

\bibitem{zeta2230}
R.M.~Baltrusaitis et al. (MARK III Collaboration),
Phys. Rev. Lett. 55 (1985) 1723.

\bibitem{gkp}
S.~Godfrey, R.~Kokoski, and J.~Paton, Phys. Lett. B54 (1985) 869.

\bibitem{phiphiMII}
D.~Bisello et al., (DM2 Collaboration), Phys. Lett. 241B (1990) 617;\\
Z.~Bai et al. (MARK III Collaboration), Phys. Rev. Lett. 65 (1990) 2507.

\bibitem{wa67}
P.S.L.~Booth et al., Nucl. Phys. B273 (1986) 677, 689.

\bibitem{jetset}
P.D.~Barnes et al., Phys. Lett. B 309 (1993) 469.

\bibitem {palchu}
A.Palano, Proc. of the Hadronic Session of the XXII Rencontre
de Moriond (1987), Ed. Fronti\`eres; \\
S.U. Chung, Z. Phys. C46, (1990) S111; \\
L. K\"opke and N. Wermes, Phys. Rep. 174 (1989) 67.

\bibitem {baipwa}
Z. Bai et al., Phys. Rev. Lett. 65 (1990) 2507.

\bibitem {coff}
D. Coffman et al., Phys. Rev. D41 (1990) 1410.

\bibitem {rhog}
T.A. Armstrong et al., Z. Phys. C54 (1992) 371.

\bibitem {ewa76}
T.A. Armstrong et al., Phys. Lett.  146B (1984) 273; \\
Z. Phys. 34C (1987) 23; \\
Phys. Lett. B221 (1989) 216; \\
Z. Phys. C56 (1992) 535.

\bibitem {pi4wa76}
T.A. Armstrong et al., Phys. Lett. B228 (1989) 536.

\bibitem {dprime}
Ph. Gavillet et al., Z. Phys. C16 (1982) 119; \\
D. Aston et al., Phys. Lett. B201 (1988) 573.

\bibitem {becker}
J.J. Becker et al., Phys. Rev. Lett. 59 (1987) 186.

\bibitem {caldwell} D.O. Caldwell, Mod. Phys. Lett. A2 (1987) 771.

\bibitem {ishida}
S. Ishida et al., Progress Theor. Phys. 82 (1989) 119.

\bibitem {longacre}
R.S. Longacre, Phys. Rev. D42 (1990) 874.

\bibitem {rhogams}
D. Alde et al., Phys. Lett. B205 (1988) 397.

\bibitem {protvino}
G.M. Beladidze et al., Phys. Lett. B313 (1993) 276.

\bibitem {gutsche}
T. Gutsche et al., Nucl. Phys. A558 (1993) 63c.

\bibitem {bauer}
D.A. Bauer et al., Phys. Rev. D48 (1993) 3976.

\bibitem {etkin}
A. Etkin et al., Phys. Lett. B201 (1988) 568.

\bibitem {omega2} D. Alde et al., Phys Lett. B205 (1988) 451; \\
G.M. Beladidze et al., Z. Phys. C54 (1992) 367.

\bibitem {rhorho} R. Brandelik et al., Phys. Lett. B97 (1980) 448; \\
H. Albrecht et al., Phys. Lett. B217 (1989) 205.

\bibitem {rhorhom} R.M. Baltrusaitis et al., Phys. Rev. D33 (1986) 1222; \\
L. Stanco, LAL 87-42, August 1987.

\bibitem {slac} See the talks of T. Bolton, C. Heusch, A. Seiden and W. Toki
in the Proceedings of the Tau-Charm Factory Workshop, SLAC, California,
23-27 May 1989. Eds. W.T. Kirk and M.L. Perl, SLAC-Report-343 (1989).

\end{thebibliography}
\end{document}